\documentclass{elsart}
\usepackage{graphicx,amssymb}
\usepackage[round]{natbib}
\journal{New Astronomy}

\makeatletter
\def\elsartstyle{%
	\def\normalsize{\@setfontsize\normalsize\@xiipt{14.5}}
	\def\small{\@setfontsize\small\@xipt{13.6}}
	\let\footnotesize=\small
	\def\large{\@setfontsize\large\@xivpt{18}}
	\def\Large{\@setfontsize\Large\@xviipt{22}}
	\skip\@mpfootins = 18\p@ \@plus 2\p@
	\normalsize
}
\makeatother

\def\astrobj#1{#1}
\def\url#1{{\ttfamily\def\/{/\discretionary{}{}{}}#1}}

\begin{document}
\bibliographystyle{abbrvnat}

\begin{frontmatter}

\title{Probing the Stellar Surface of HD209458 from Multicolor Transit
Observations}



\author[iaa,iac]{H.J. Deeg\thanksref{emailhj}}
\author[iaa]{R. Garrido}
\author[iaa]{A. Claret}

\thanks[emailhj]{E-mail: users \emph{hdeeg, garrido, claret} at
\emph{user@iaa.es}}

\address[iaa]{Instituto de Astrof\'\i sica de Andaluc\'\i a, C. Bajo 
de Hu\'{e}tor 24, Granada, Spain}
\address[iac]{Instituto de Astrof\'\i sica de Canarias, C. V\'\i a
L\'{a}ctea, La Laguna, Tenerife, Spain}

\begin{abstract}
Multicolor photometric observations of a planetary transit in the
system \astrobj{HD~209458} are analyzed. The observations, made in the
Str\"omgren photometric system, allowed a recalculation of the basic
physical properties of the star-planet system. This includes
derivation of linear limb-darkening values of \astrobj{HD~209458},
which is the first time that a limb-darkening sequence has observationally
been determined for a star other than the Sun. As the derived physical
properties depend on assumptions that are currently known with limited
precision only, scaling relations between derived parameters and
assumptions are given. The observed limb-darkening is in good
agreement with theoretical predictions from evolutionary stellar
models combined with ATLAS model atmospheres, verifying these models
for the temperature ($T_{eff} \approx 6000K$), surface gravity ($\log
g \approx 4.3$) and mass ($\approx 1.2 M_\odot$) of HD~209458.
\end{abstract}

\end{frontmatter}

\section{Introduction}
\label{intro}
The discovery of the first extrasolar planet transiting its central
star, \astrobj{HD~209458}
\citep{2000ApJ...529L..45C,2000ApJ...529L..41H} opened a new way to
study these objects through the analysis of the light curve that is
generated during a transit. Similarly, details about the central star
may be revealed. Here we present observations of a transit of
\astrobj{HD~209458} that were taken simultaneously in the four
Str\"{o}mgren colors. The primary motivation was a study of the
specific intensity distribution of the stellar surface of
\astrobj{HD~209458} for comparison with existing models of stellar
atmospheres through the derivation of limb-darkening coefficients. A
transit light curve provides in its central part -- between second and
third contact -- a direct measure of the specific intensity behind the
planet. The specific intensity behind the ingress and egress zones is
also accessible, through models of transit light curves. A planetary
transit profile gives significant advantages against efforts to derive
intensities across the stellar disc from light curves of eclipsing
binaries: in transits, a small dark disk passes in front of the star,
whereas eclipsing binaries cause ambiguities from the presence of two
luminous sources, frequently of similar size range, and often
aggravated by light reflection between the two components, or by
gravitational darkening in the case of fast rotators.

The major difference between the data presented here and the $BVRIZ$
multicolor transit observations by \citet{2000ApJ...540L..45J} is the
different wavelength span, with our data taken in the
Str\"{o}mgren photometric system. Also, our data were taken in all
colors simultaneously, whereas \citeauthor{2000ApJ...540L..45J} used a
CCD camera cycling sequentially through the $V$, $R$ and $I$ filters,
with data in further colors obtained at telescopes in different
locations. In the following analysis we then show, that transit light
curves do not only allow the derivation of parameters of the
extrasolar planet, but impose severe constraints on the mass-to-radius
relationship of the central star as well. Furthermore, derived
parameters of the star-planet system are presented in a way that is
independent of assumptions based on approximately known values, such
as the stellar mass. Finally, linear limb-darkening coefficients are
given as a function of the stellar radius and mass of
\astrobj{HD~209458} and compared to stellar and atmosphere
models. This is the first time that a color sequence of limb-darkening coefficients for a
single star besides the Sun has been derived from observations.

\begin{figure}
\begin{center}
\includegraphics*[height=18cm]{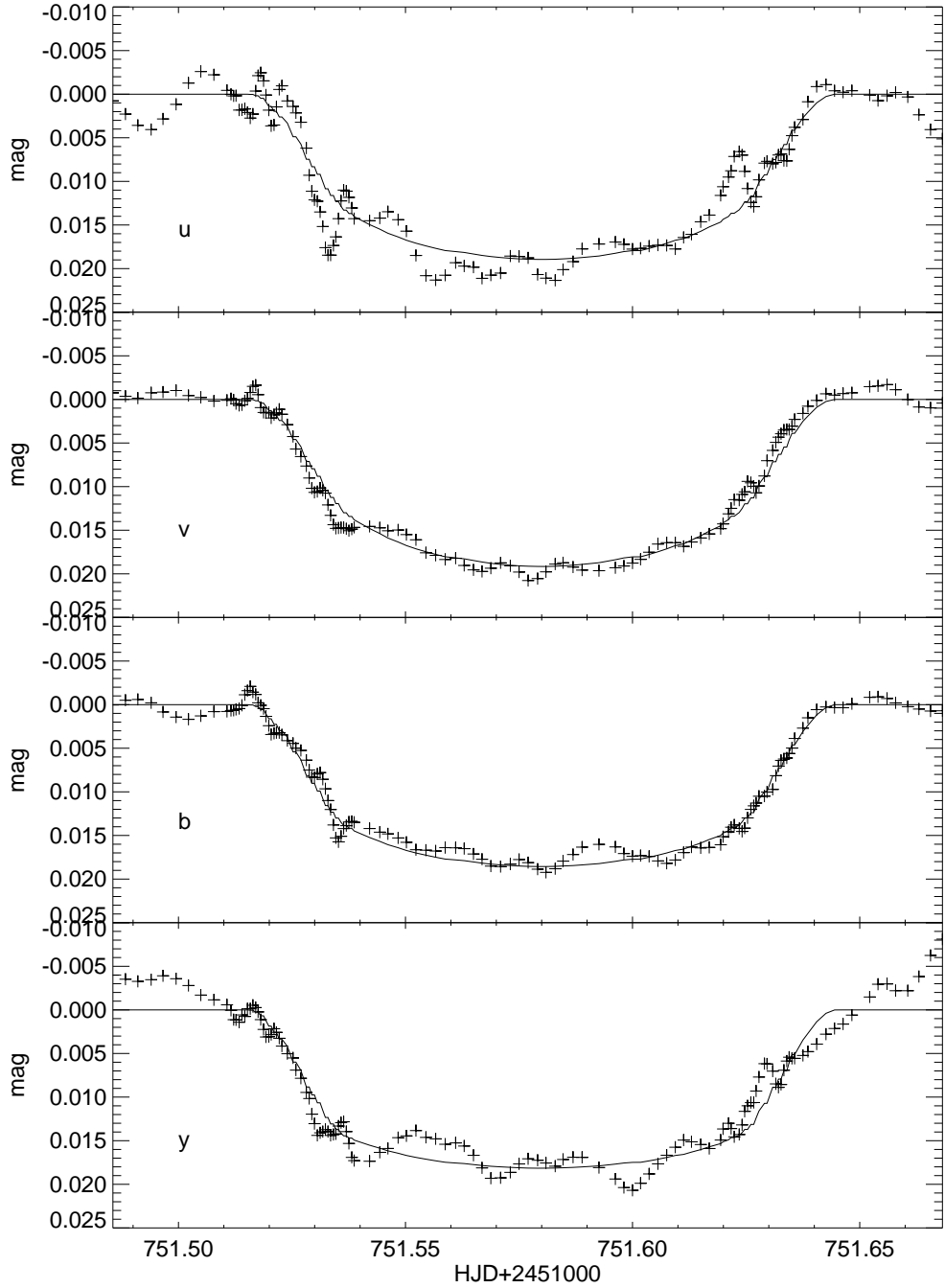}
\end{center}
\caption{$u,v,b,$ and $y$ light curves that were used as the basis for
the analysis (crosses). They are smoothed by a triangular kernel with
a width of 5 data points. The solid lines are model light curves for a
star-planet system with the parameters given in Table~\ref{tab:params}.}
\label{fig:4lc}
\end{figure}

\section{Data acquisition}
\label{obs}
The observations of \astrobj{HD~209458} reported here were performed
in the night of 25-26 July 2000 with the 0.9m telescope of the
Observatorio Sierra Nevada, located near Granada, Spain at an altitude
of 2900m. The telescope is equipped with a 4 channel photometer,
allowing simultaneous observations in all colors of the $u,v,b,y$
Str\"{o}mgren photometric system.  This instrument is similar to that
of the Str\"{o}mgren Automatic Telescope at La Silla
\citep{1976A&AS...26..155G}. The observing cycle consisted of a 30 sec
sky-exposure, 20 sec exposures of standard stars SAO~107675 and
SAO~107561, and a 40 sec exposure of the object,
\astrobj{HD~209458}. During the ingress and egress of the transit, for
about 40 minutes each time, only \astrobj{HD~209458} was observed with
repeated 40 second exposures. During the flat bottom part of the
transit, the second comparison star was left out. These modifications
served to increase the time resolution, and were justifiable since the
night was photometric, with last quarter moon, and the mid-transit
occurred within a few minutes of the meridian crossing of
\astrobj{HD~209458}, leading to equal airmasses at ingress and
egress. Data reduction was performed at the telescope Macintosh
control computer, with the derivation of an extinction function based
on the comparison stars. Differential photometry was then performed
with respect to the first comparison star. Resulting lightcurves are
available in electronic Table~1.  \addtocounter{table}{1} Magnitude
differences between both comparison stars indicate a photometric
precision of about 4 mmag for the $u$ and $y$ bands and 2 mmag for the
$v$ and $b$ ones per single measurement, with the better results in
the two latter bands simply being due to higher photon count
rates. After removal of poorer data at the begin of the observations,
the transit light curves used in the analysis contain 129 data points
in all colors, with 29 off-transit points, which were used to set the
magnitude zero-point\footnote{For the $y$ filter, shorter term
variations not well described by the extinction function required
offsets from the off-transit zero-point found through adjustment of
the best fitting planet size (see Fig.~\ref{fig:rplbest}) to that of
the other colors}. In the off-transit part, the rms noise is: 4.2mmag
for $u$, 2.2mmag for $v$, 1.9mmag for $b$ and 3.4mmag for $y$. For the
subsequent analysis, the light curves were smoothed with a triangular
kernel of 5 elements width, obtained through two successive boxcar
smoothings with a 3 element rectangular kernel. The off-transit rms of
the resulting light curves, shown in Fig.~\ref{fig:4lc}, is then: $u$:
1.8mmag, $v$: 0.72mmag, $b$: 0.89mmag, $y$: 2.5mmag. Some parts of our
analysis are based on a combined $v+b$ light curve, for which a lower
rms of 0.65mmag was attained. Similar observations as reported here
were performed in the night of July 18-19, 2000 as well, but
increasingly hazy weather conditions and a nearly full moon
significantly degraded these observations, leading us to concentrate
on the data from 26 July.

\section{Analysis of the transit}
\label{params}
\subsection{Setup of model fits to the transit} 
 
A convenient set of 'primary transit parameters', from which
other frequently used parameters (like the planet's orbital
inclination $i$ or the orbital half axis) can easily be derived,
may be selected as follows:

$P$: orbital period\\ $T_c$: epoch of the planetary transit\\ $M_*$:
stellar mass\\ $R_*$ : stellar radius\\ $R_{\rm{pl}}$:
\emph{effective} planetary radius, which is the radius of an opaque
disk occulting the same amount of light as the planet.\\ $b$: latitude
of the transit across the stellar disk in stellar surface
coordinates\\ 
$w$: linear limb darkening coefficient.

$w$ is representative for some descriptor of the intensity
distribution across the stellar disk. Here, we will use the common
linear limb-darkening expression, defined through $I(\cos
\theta)=I(0)\ [1-w(1-\cos \theta)]$, where $\theta$ is the angle
between the stellar surface and the line of sight. The use of a linear
limb-darkening expression for \astrobj{HD~209458} is justified, since it gives a
good description of surface intensities for stars with effective
temperatures between 5000 and 6000~K, and $T_{eff}$~=~6000~K for
\astrobj{HD~209458} \citep{2000ApJ...532L..55M}.

Among the primary parameters, only the orbital period and the epoch
are known with high precision
\citep{1999IBVS.4816....1S,2000ApJ...532L..51C,2000A&A...355..295R}.
For this work, a period of 3.52474 days and an epoch of HJD
2451659.93675 from HST observations by \citet{bcg+2001} was used, and -- upon excellent
agreement with our data -- no attempt for a new derivation has been
undertaken. Furthermore, estimates with a precision of about 10\%
exist for $R_*$ and $M_*$, with $R_* = 1.2 R_\odot$ and $M_* =
1.1 R_\odot$ \citep[ and references therein]{2000ApJ...532L..55M}.

A boundary condition that allows a reduction in the number of free
parameters is given by the total length of the transit, which is the
time $T_{14}$ between the first and the last -- fourth -- contact of star
and transiting planet. The observations by
\citet{bcg+2001} give a value for $T_{14}$ of 0.12795 days, which is
in excellent agreement with our data. $T_{14}$ is given by the
following expression, whose derivation is shown in Fig.~\ref{fig:t14}:

\begin{figure}[tb]
\begin{center}
\includegraphics*[]{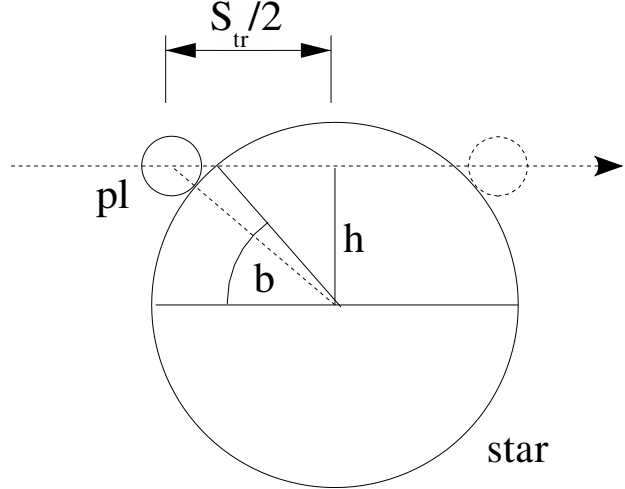}
\end{center}
\caption{The total transit duration $T_{14} = S_{\rm{tr}}
/v_{\rm{orb}}$ depends on the lateral length of the transit 
$S_{\rm{tr}}$ 
as follows: $(S_{\rm{tr}}/2)^2 +
h^2 = (R_{\rm{pl}} + R_*)^2$, with $R_{\rm{pl}}$ and $R_*$ being the
radii of the planet and of the star. If $h \le R_*$, then
$h$ may be expressed in terms of the latitude $b$ of the transit
across the surface of the star: $h = R_* \sin b$, and consequently,
$h^2 = R_*^2 (1 - \cos^2 b)$.  Solving this for $S_{\rm{tr}}$ leads
then to Eq.(1).}
\label{fig:t14}
\end{figure}

\begin{equation}
T_{14} = \frac{2 R_{*}}{v_{\rm{orb}}(M_{*},P)} \left[\cos ^{2}b 
+\left(\frac{R_{\rm{pl}}}{R_{*}}\right)^{2} + \frac{ 2 
R_{\rm{pl}}}{R_{*}} 
\right]^{1/2}
\end{equation}

where the planet's orbital velocity $v_{\rm{orb}}$ depends in 
circular orbits\footnote{For the case
of \astrobj{HD~209458}b -- a planet that is on a very close orbit around its
central star -- the actual $T_{14}$ will be, in the worst case, 1.0024
times longer than derived by Eq.(1)} on $M_*$ and $P$ as $v_{\rm{orb}}
\propto (M_* /P) ^{1/3}$. With $T_{14}$ known, Eq.(1) can be solved 
for
$\cos b$ as a function of $R_*, M_*$ and $R_{\rm{pl}}$:

\begin{equation}
\cos b = \left[ \left( \frac{T_{14} \ v_{\rm{orb}}} {2 R_{*}} \right)
^{2} - \left( \frac{R_{\rm{pl}}}{R_{*}} \right)^{2} - \frac{2
R_{\rm{pl}}}{R_{*}} \right] ^{1/2}
\end{equation}

which corresponds to physically meaningful solutions only for $0 \le
\cos b \le 1$.

For a given combination of the parameters $\cos b, R_*, R_{\rm{pl}}$,
an estimate for the limb-darkening coefficient $w$ can then be
obtained as follows:\\ The relative brightness change $\Delta L/L$ at
the center of a transit, assuming the linear limb-darkening expression, is
given by:

\begin{equation}
\Delta L/L = \left( \frac{R_{\rm{pl}}}{R_*} \right)^2\ 
\frac{3}{(3-w)}\ \left[1-w(1-\cos b)\right]
\end{equation}

which can be derived from the linear limb-darkening equation, with
$\theta=b$ at the center of a transit. With an estimate of $\Delta
L/L$ taken from the observed light curve, the above equation can be
solved for $w$:
\begin{equation}
w = \frac{3\left[(R_{\rm{pl}}/R_*)^2 - \Delta L /L 
\right]}{3(R_{\rm{pl}}/R_*)^2(1-\cos b) - \Delta L/L} 
\end{equation}

\begin{figure}[tb]
\begin{center}
\includegraphics*[width=14cm]{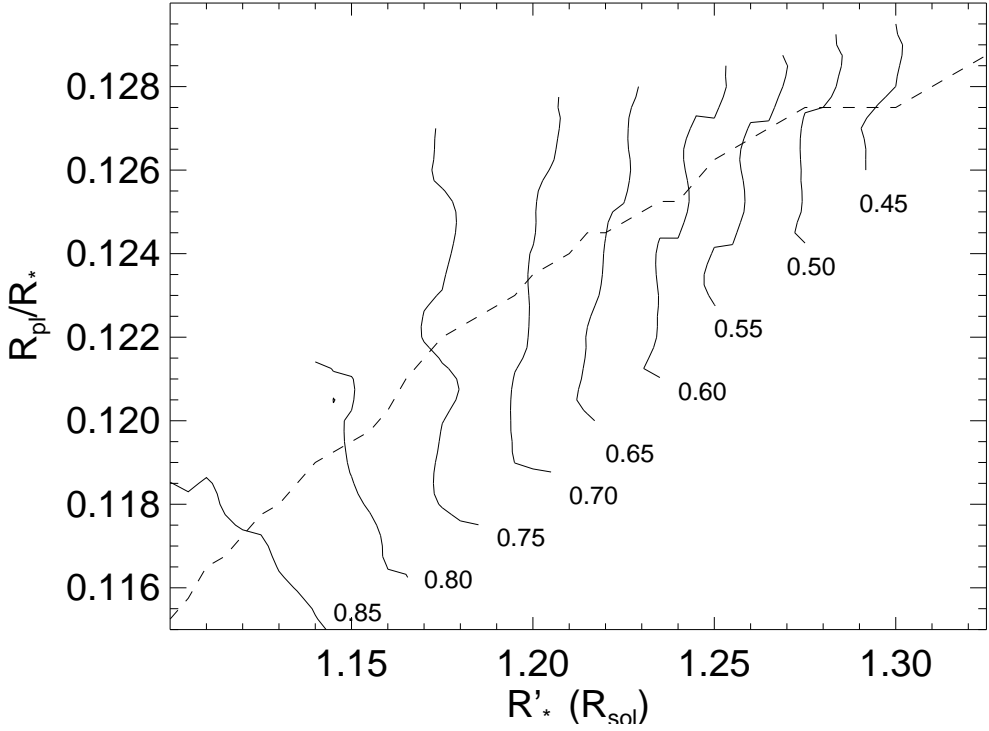}
\end{center}
\caption{The contours give limb darkening values $'w'$ for the
combined $v+b$ color, as a function of the boundary parameters
$R_{\rm{pl}}/R_*$ and $R_*$, for a fixed stellar mass of 1.1
$M_{\odot}$ (or $R_{\rm{pl}}/R_*$ and $R'_*$ for any stellar mass, see Sect.~\ref{starpar}). The limited vertical lengths of the contours is given by
the range in which solutions where calculated.  The dashed line gives
the trace of the best fit for each value of $R'_*$ (see also
Fig.~\ref{fig:rplvsrs}).  Small changes in the planet size (up-down)
around the best fit have little effect on $w$.}
\label{fig:uvsrs}
\end{figure}

\begin{figure}[tb]
\begin{center}
\includegraphics*[width=14cm]{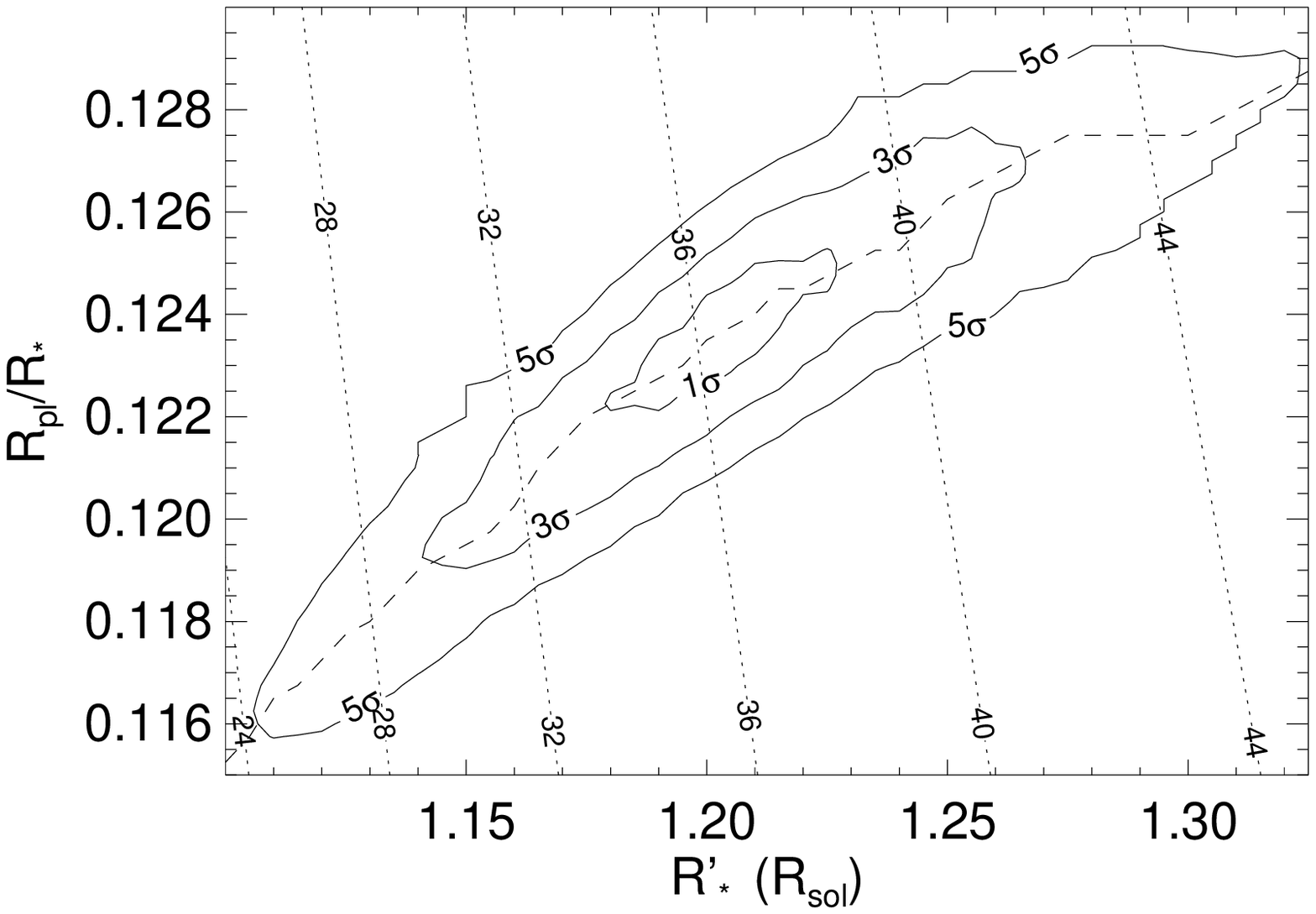}
\end{center}
\caption{Confidence regions (1, 3 and 5 $\sigma$) of the best fit
to $w$, within the same parameter space as the previous figure. The
dashed line traces the best fitting value of $R_{\rm{pl}}/R_*$ for a
given $R'_*$. Dotted lines indicate the latitude (in degrees) of the 
transit, from Eq.(2).
}
\label{fig:rplvsrs}
\end{figure}

Fits to the light curve were performed using the transit modeling code
UTM, which calculates light curves following a set of input
parameters, and the fitting routine 'UFIT'\footnote{UTM (Universal
Transit Modeler) and UFIT (Universal Fitter) are written in IDL v.5
code and are available from HJD at
\url{http://www.iac.es/galeria/hdeeg/idl\_hans\_lib/utm/}. UTM can be
used in stand-alone mode to produce transit light curves, or within
the fitting routine UFIT. UTM calculates the light curves of transits
from pixelized representations of projections of the stellar surface
and the transiting object. For both UFIT and UTM, setup files are used
to completely describe a transit configuration, and allow the
representation of systems with several stars and planets, moons or
rings. UFIT can perform fits to any of the input parameters of UTM.},
which wraps around UTM and provides UTM with input parameters
following a Levenberg-Marquardt fitting algorithm. UFIT then stepped
through a 3 dimensional grid of boundary values for $M_*$, $R_*$ and
$R_{\rm{pl}}/R_*$, performing fits of the only free parameter $w$ at
each step. The boundary values were varied within the following
limits:

$0.95 M_\odot \le M_* \le 1.35 M_\odot$, with increments of $0.025 
M_\odot$\\
$0.95 R_\odot \le R_* \le 1.35 R_\odot$, with increments of $0.025 
R_\odot$\\
$0.11 \le R_{\rm{pl}}/R_* \le 0.145$, with increments of 0.00125

At each step, a new value for $\cos b$ was calculated from Eq.(2).  If
a physically meaningful solution for $\cos b$ was found, an initial
value for $w$ was calculated from Eq.(4). Only if this value was in
the range of $0.2 \le w_{\rm{init}} < 1$, an actual fit to the light
curve was performed.  Since $w$ does not appear in Eq.(1), results
from the fits will always comply with the boundary condition of the
total transit duration $T_{14}$.

The result from these 'scanning fits' are two three-dimensional arrays
with the axes along $M_*$, $R_*$ and $R_{\rm{pl}}/R_*$, one containing
values for the fitted limb-darkening coefficient $w$ and the other one
the variances between fit and data.  A 'cut' through such an array of
$w$ values, from fits to the combined $v+b$ light curve, is shown in
Fig.~\ref{fig:uvsrs}, and the corresponding variances, converted to
confidence regions, are shown in Fig.~\ref{fig:rplvsrs}. All 
confidences in this paper are calculated for 129 data points and 4 
free parameters.

\subsection{Dependence on stellar mass and 
radius} 
\label{starpar}

\begin{figure}[tb]
\begin{center}
\includegraphics*[width=14cm]{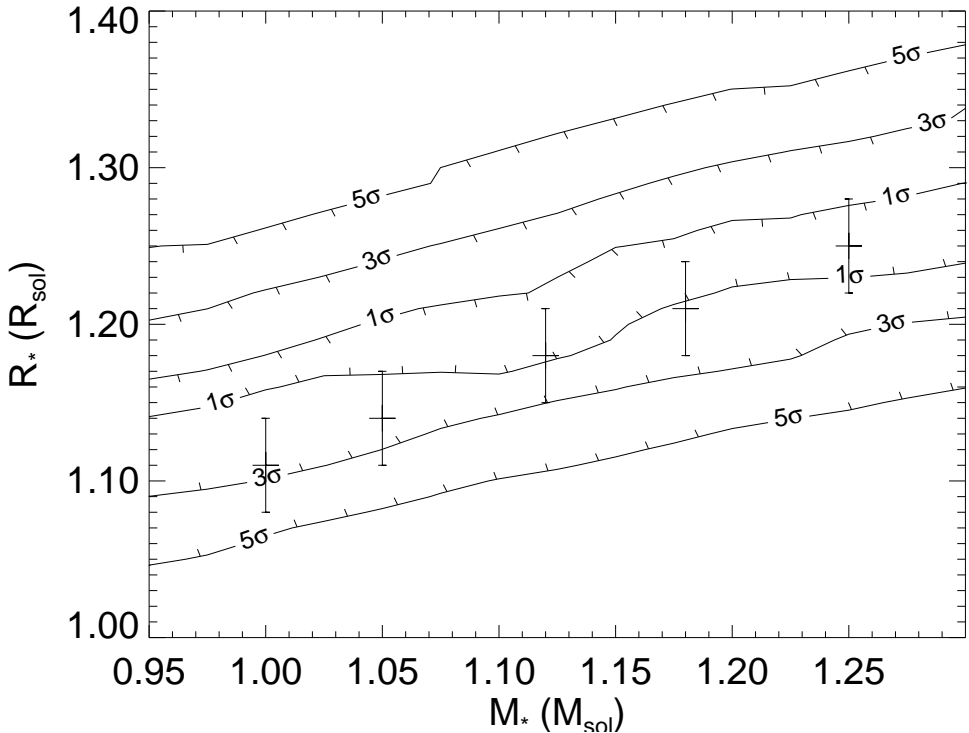}
\end{center}
\caption{1, 3, and 5 $\sigma$ confidence regions of the best fit on 
the combined $v$ and $b$ light curve, as a function of the mass and 
radius of the central star. Best fits lie along a band in 
$(M_*,R_*)$, also given by Eq.(5). The 5 points with error bars 
are 
those radii where limb-darkening coefficients from  
transit data and from ATLAS atmosphere models coincide (see 
Section~\ref{limb} and Fig.~\ref{fig:limb_claret}).}
\label{fig:M_vs_R}
\end{figure}

In the following, we will use the light curve with the lowest
noise -the combined one from $v$ and $b$ filters- for the derivation
of all parameters, except for the limb-darkening coefficients, which
are wavelength dependent. In Fig.~\ref{fig:M_vs_R}, the 3-dimensional
array that resulted from the scanning fit is flattened into 2
dimensions ($M_*,R_*$) by selecting those values of $R_{\rm{pl}}/R_*$
which gave the best fit at each point $M_*,R_*$. The best solution
appears in a narrow band, which can be approximated by

\begin{equation}
R_* = 0.34 M_* + 0.825(\pm 0.06) , 
\end{equation}

with the errors indicating a $3 \sigma$ confidence limit. This relationship follows lines of constant transit 
latitude, which is dominated by $\cos b 
\propto 
v_{\rm{orb}}/R_{*} \propto M_*^{1/3}/R_{*}$ for $R_{\rm{pl}} << R_{*}$ 
(see Eq.2). It should be noted that this 
relationship has no physical meaning in the sense of compatibility 
with stellar models. It is simply a result from the transit light 
curve and the $T_{14}$ boundary condition, giving the range of possible radii as a function of the stellar 
mass that are compatible with transit observations. Eq.(5) allows us to 
express most results given in the further sections as functions of a 
'normalized radius' $R'_*$, which is related to the actual stellar mass 
and radius (currently known with large errors only) as follows:

\begin{equation}
R'_* = R_* + 0.34 (1.1- M_*)
\end{equation}

For $M_*= 1.1$, normalized and actual radii are equal. It should be
noted that the stellar parameters given by \citeauthor{2000ApJ...532L..55M} ($M_* = 1.1
M_{\odot}, R_*= 1.2 R_{\odot}$) lie well within the 1$\sigma$
confidence region of Fig.~\ref{fig:M_vs_R}. The radius of $1.27\pm
0.05 R_{\odot}$ from transit observations by
\citet{2000ApJ...540L..45J} is slightly outside, and the value of
$1.146\pm 0.050 R_{\odot}$ from HST observations
(\citeauthor{bcg+2001}) is slightly inside our 3$\sigma$ confidence
region (both times assuming a stellar mass of $1.1 M_{\odot}$), with
error bars from the HST observations well overlapping with our
determination.

\subsection{Planet size, transit latitude and inclination}
\label{plansize}

\begin{table*}[tb]
\begin{center}
\caption{\label{tab:plansize} Determinations of planet and star sizes
in the \astrobj{HD~209458} system}
\begin{tabular}{lccc} 
\hline
Reference&$R_{\rm{pl}} (R_{\rm{Jup}})$&$R_* (R_\odot)$&$R_{\rm{pl}}/R_*$\\
\hline
\citealp{2000ApJ...529L..45C}&$1.27\pm0.02$&1.10&$0.1193\pm0.002$ \\
\citealp{2000ApJ...529L..41H}&$1.42\pm0.10 $&1.15&$0.1276\pm0.009$ \\
\citealp{2000ApJ...532L..55M}&$1.40\pm0.17 $&1.20&$0.1205\pm0.015$ \\
\citealp{2000ApJ...540L..45J}&$1.55\pm0.10$&1.27&$0.1261\pm0.008$ \\
\citealp{bcg+2001}&$1.347\pm0.06$&1.146&$0.1214\pm0.002$\\
this work&$1.435\pm0.05$&1.20&$0.1235\pm0.004$\\
\hline
\end{tabular}
\end{center} 
\vspace{0.5cm}
\end{table*}
 
In Fig.~\ref{fig:rplvsrs}, the confidence regions along
$R_{\rm{pl}}/R_*$ and $R'_*$ are shown. It should be noted, that the
relative planet size $R_{\rm{pl}}/R_*$ is a function of $R'_{*}$,
whereas the absolute size $R_{\rm{pl}}$ is \emph{not}. The region of fits with
any significant meaning ($\sigma < 3 \sigma_{\rm{min}}$) is limited to
a range of stellar and planetary radii given by: $R'_* = 1.20 \pm 0.06
R_{\odot}$ and $R_{\rm{pl}}/R_* = 0.1235 \pm 0.004$.  For a stellar
radius of $1.2R_\odot$, the derived planet size would then be
$1.435\pm0.05 R_{\rm{Jup}}$. There is considerable scatter in the
literature among derivations for the absolute size of the planet
(Table~\ref{tab:plansize}), which is mainly caused by differences in
the assumed stellar size.  The agreement is much better if relative
planet sizes are considered, whose average from all listed
publications is $R_{\rm{pl}}/R_* = 0.1231 \pm 0.003$, which is very
close to the value derived in this work.

\begin{figure}[tb]
\begin{center}
\includegraphics*[width=14cm]{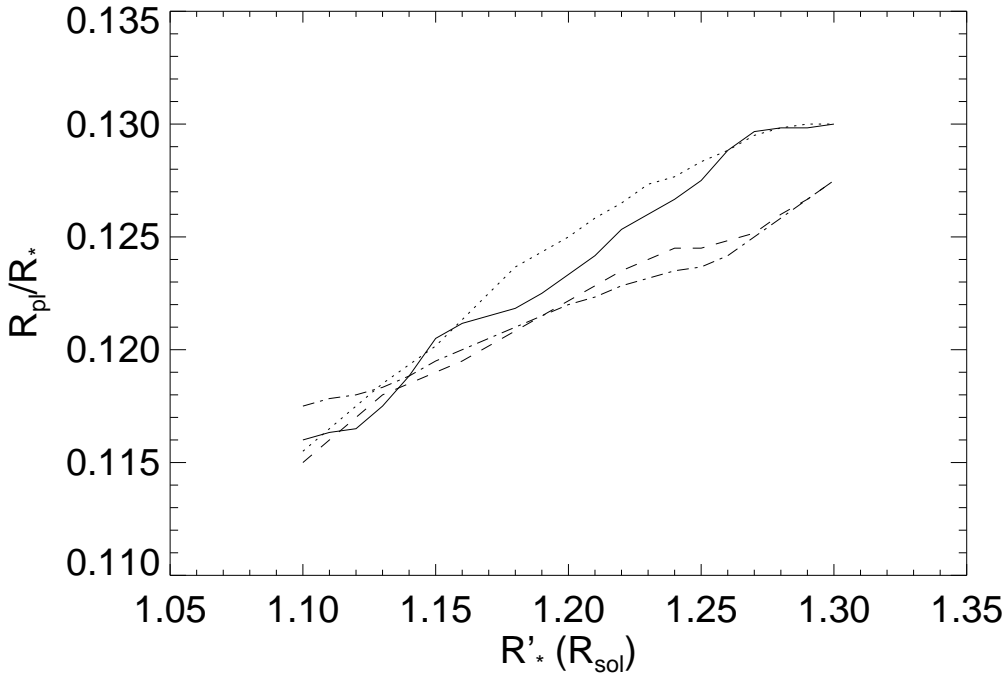}
\end{center}
\caption{Best fitting planet sizes in the four Str\"{o}mgren colors,
as a function of the stellar size ($u$: solid line; $v$: dotted; $b$:
dashed; $y$: dot-dashed) }
\label{fig:rplbest}
\end{figure}

The best fitting planet sizes in each of the four Str\"omgren colors
are shown in Fig.~\ref{fig:rplbest}. Their scatter around the mean is
better than $\pm$ 3 \%. This shows that the depths of the transit
across the different colors agree with each other, and that the
magnitude zero points in the light curves were correctly set. The
remaining differences among the individual light curves will then
primarily be caused by the dependence of limb-darkening on the
wavelength.

Independently of the stellar mass, the range of possible transit
latitudes is restricted to $ b = 36.2 \pm 6^\circ$
(Fig.~\ref{fig:rplvsrs}, within the 3 $\sigma$ confidence region). For
$M_* = 1.1 M_{\odot}$, this corresponds to an inclination of the
planet's orbit of $i = 85.96\pm 0.55^\circ$, given by:
\begin{equation}
\cos i = \frac{R_{*}}{a_{\rm{pl}}} \sin b
\end{equation}
where the planetary orbit's half axis $a_{\rm{pl}}(M_*, M_{\rm{pl}},
P)$ is given by Kepler's Third Law ($a_{\rm{pl}} = 10.057 R_\odot$ for
$M_* = 1.1 M_{\odot}$).

\begin{figure}[tb]
\begin{center}
\includegraphics*[width=14cm]{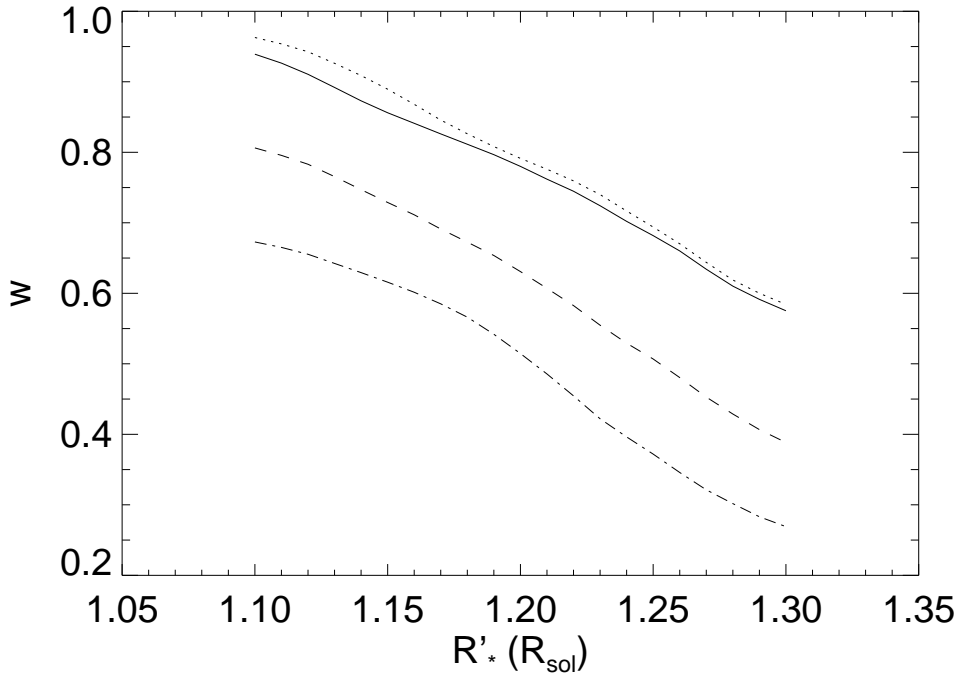}
\end{center}
\caption{Linear limb-darkening coefficients for \astrobj{HD~209458} in the
Str\"{o}mgren colors ($u$: solid line; $v$: dotted; $b$: dashed; $y$:
dot-dashed) as derived from the fits. They are plotted against the
normalized stellar radius of Eq.(6) and correspond to traces along
best fitting values of $R_{\rm{pl}}/R_*$, similar to the dashed line
in Fig.~\ref{fig:uvsrs}.} \label{fig:limb}
\end{figure}

\section{The limb-darkening of \astrobj{HD~209458}}
\label{limb}
The excellent fit between the modeled light curves and data 
(Fig.~\ref{fig:4lc}) shows that the use of a linear limb-darkening expression 
has been adequate for this star.

Values of the limb-darkening are shown for the four Str\"{o}mgren
colors in Fig.~\ref{fig:limb}, again quoting $w$ for the best fit at
each value of $R'_{*}$, similar to the dotted line in
Fig.~\ref{fig:uvsrs}.  Fig.~\ref{fig:uvsrs} also shows that the
derived values of $w$ are not very dependent on small deviations of
$R_{\rm{pl}}$ around its best value. Individual errors of the
limb-darkening coefficients are estimated at $\pm 0.05$ for $w_v$ and
$w_b$, $\pm 0.07$ for $w_y$ and $\pm 0.1$ for $w_u$.  The derived
values agree reasonably well with the usual sequence of limb-darkening
coefficients along the four colors, with a distance between $w_u$ and
$w_v$ that is smaller than their uncertainty. It is also apparent,
that only for a small range of stellar radii, reasonable
limb-darkening values can be assigned to \astrobj{HD~209458}. For two
narrow wavelength ranges, centered around 590nm and 618nm,
\citet{bcg+2001} provide quadratic limb-darkening values, which
correspond to linear coefficients of $w_{590} \approx 0.62$ and
$w_{618} \approx 0.58$. These wavelength ranges are slightly red-wards
of the Str\"{o}mgren $y$ filter ($\lambda = 547$nm), and their
limb-darkening agrees very well $w_y \approx 0.62$ from our data,
assuming a stellar radius of 1.146 $R_\odot$ as did
\citeauthor{bcg+2001}.

\begin{figure}[tb]
\begin{center}
\includegraphics*[width=14cm]{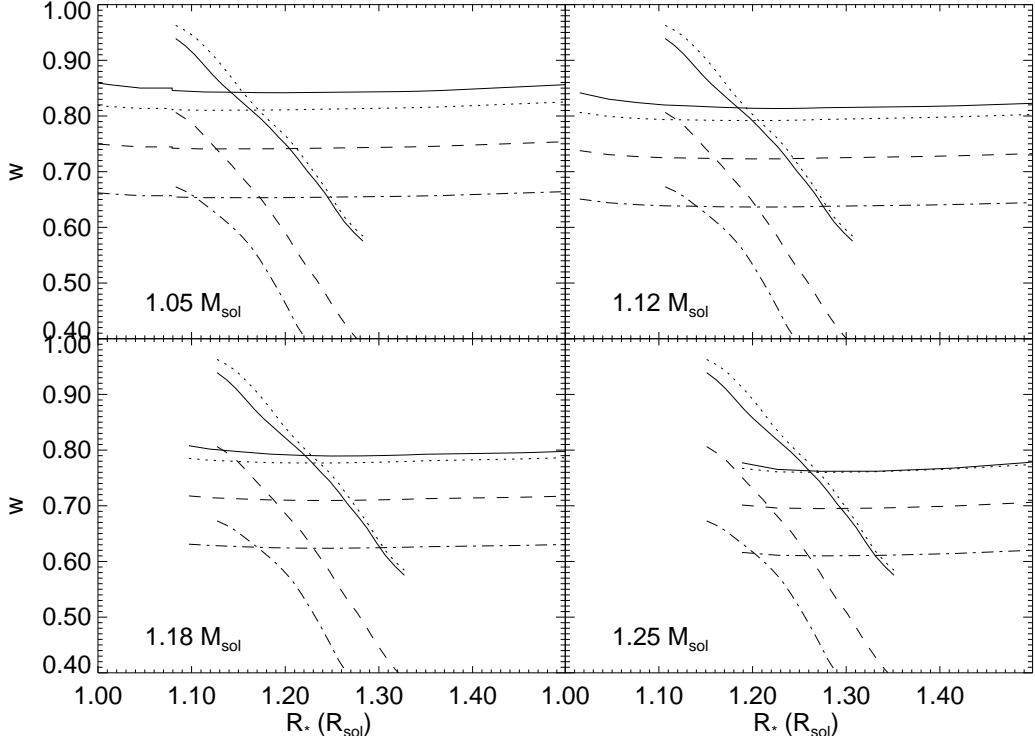}
\end{center}
\caption{Limb darkening coefficients derived from the transit data
(inclined curves) against values based on the ATLAS model atmosphere
combined with evolutionary stellar models \citep[flat
curves]{1998A&AS..131..395C}. Plots are shown for stellar masses of
1.05 - 1.25 $M_{\odot}$, with the normalized radii ($R'_*$) of
Fig.~\ref{fig:limb} converted to $R_*$ using Eq.(6). Str\"{o}mgren
filters correspond to line-styles as in the previous figures.}
\label{fig:limb_claret}
\end{figure}

The limb-darkening coefficients have been compared with values
derived from the combination of evolutionary stellar models
\citep{1995A&AS..109..441C} and stellar atmosphere calculations based
on the ATLAS code, using tabulations of $w$ from
\citet{1998A&AS..131..395C}.  All the models have solar abundances and
surface gravity values were converted directly to stellar
radii.  Comparison with limb-darkening coefficients derived from
PHOENIX code \citep{1998A&A...335..647C} was not possible because the
code does not cover the temperature range of \astrobj{HD~209458}.  The results
from the evolutionary ATLAS models are shown in Fig.~\ref{fig:limb_claret}. Although
a good general agreement between the limb-darkening sequence from
transits and from ATLAS is found, the values from the transit have a
wider spread across the four Str\"omgren colors. This can be seen
through the intersections among the curves of $w$, which do not
correspond to exactly the same stellar radii across the four
colors. However, the intersections appear close to those stellar radii
at which the best fits have been obtained. These intersection radii,
averaged among the colors, are displayed in
Fig.~\ref{fig:M_vs_R}(crosses) and have a somewhat steeper dependency
on $M_*$ than the radii directly derived from transits. For $M_* >
1.25 M_\odot$, no stellar model delivers radii small enough for those
derived from transits, and for masses less than 1.05 $M_{\odot}$,
radii from ATLAS correspond to transit-radii outside of the $2\sigma$
confidence region.  The radii based on limb-darkening provide thereby
a constraint for the stellar mass to be within 1.05 and 1.25
$M_{\odot}$, with a stellar mass of $1.1M_\odot$ leading to a radius
of $1.17\pm0.03 R_\odot$.

The standard linear limb-darkening expression appears to be sufficient
to account for our observations.  For one, \astrobj{HD~209458} is in the
temperature range where the linear expression is expected to give good
results.  Second, most of the contributing non-linearity would come
from angles close to the stellar rim, which are within the ingress and
egress zone of the transit. Parameters describing non-linearities are
therefore susceptible to introduce ambiguities with other transit
parameters. Lastly, the number of free parameters in fits should be
kept small in order to avoid the derivation of highly interdependent
parameters of low significance. For stars outside of the temperature
range of 5000-6000 K, single-parametric limb-darkening expressions
different to the one used here may be devised from modeled surface
brightness distributions. An example that this can be done with a very
high precision, describing limb-darkening of the Sun, is given by
\citet{1998A&A...333..338H}.

A possible planetary atmosphere may also cause differential color
effects to become apparent in transit light curves. A color dependency
of the atmospheric transparency, albedo, or scale height will cause
the planet to display an effective radius that changes with colors. We
however do not expect to detect such color effects within the
precision of our data. Assuming a precision of 0.5~mmag, then $\delta
L \approx \delta A_{\rm{pl}}/{A_*}$, where $A_* = \pi R_*^2$ is the
cross section of the star, and $\delta A_{\rm{pl}}$ is the
differential cross section of the planet. With $\delta A_{\rm{pl}}$
depending on a variation of the planetary radius like: $\delta
A_{\rm{pl}} = 2 \pi R_{\rm{pl}} \delta R_{\rm{pl}}$, we may solve for
$\delta R_{\rm{pl}}$ as follows: $\delta R_{\rm{pl}} = \delta
A_{\rm{pl}} / 2 \pi R_{\rm{pl}} = R_*^2 \delta L /2 R_{\rm{pl}}$. For
$ R_{\rm{pl}} = 1.43 R_{\rm{Jup}} = 0.148R_{\odot}$ and $R_*= 1.20
R_{\odot}$ we obtain $\delta R_{\rm{pl}} = 0.0024 R_{\odot} \approx
1700 km $, meaning that color variations induced in the effective
radius need to be larger than that value to be detectable in our
data. Unless \astrobj{HD~209458} has an extremely extended atmosphere
with a very unlikely transparency dependence along the Str\"{o}mgren
colors, our data will not contain any signatures from the planetary
atmosphere. Data with significantly higher precision may however very
well indicate the presence of planetary atmospheres.

\section{Conclusions}

\begin{table*}[tb]
\begin{center}
\caption{\label{tab:params} Derived parameters for the \astrobj{HD~209458} 
system, and their interdependencies}
\begin{tabular}{lccccc} 
\hline
Parameter&Value&Assumption&Reference\\
\hline
$R_*$&$1.20 \pm 0.02 R_{\odot}$&$M_* = 1.1M_{\odot}$&Fig. \ref{fig:M_vs_R},
Eq.(5)\\
$R_{\rm{pl}}/R_*$&$0.1235\pm0.001$&$R'_*=1.20\pm0.02R_{\odot}$&Figs. \ref{fig:rplvsrs}, \ref{fig:rplbest}\\
$R_{\rm{pl}}/R_{\rm{Jup}}$&$1.435\pm 0.05$&$M_*=1.1M_{\odot},R_*=1.20\pm0.02R_{\odot}$&Figs. \ref{fig:rplvsrs}, \ref{fig:rplbest}\\
$b $&$ 36.2\pm2.2^\circ$&$R_{\rm{pl}}/R_*$ and $R'_*$ from above&Eq.(2), Fig. \ref{fig:rplvsrs}\\
$i $&$85.96\pm0.22^\circ$&$b$ and $M_*$ from above&Eq.(7)\\ 
$w_u$&$0.82 \pm 0.10$&$R'_* = 1.17 R_{\odot}$& Fig.~\ref{fig:limb}\\
$w_v$&$0.84 \pm 0.05$& `` & ``\\
$w_b$&$0.69 \pm 0.05$& `` & ``\\
$w_y$&$0.59 \pm 0.07$& `` & ``\\
\hline
\end{tabular}
\end{center} 
\vspace{0.5cm}
\end{table*}

A major goal of this work has been the derivation of limb-darkening
coefficients for the star \astrobj{HD~209458}. These, and other
parameters of the system, have been obtained in a way which is
quite independent of the underlying physical assumptions. An overview
of the derived parameters and assumptions is given in Table~\ref{tab:params}. References in the table are given to understand the interdependence
of the derived parameters upon the assumed values. Any improvement in
any parameter could be easily used, through the references given in
the table, to adjust any other parameter for the \astrobj{HD~209458}
system. For example, if we assume the stellar size of 1.146 $R_\odot$
from HST observations by \citeauthor{bcg+2001}, and a mass of
$1.1M_\odot$, the corresponding planet size, transit latitude, and
orbital inclination derived from our data would be $R_{\rm{pl}} =
1.323 R_{\rm{Jup}}, b = 30.0^\circ$ and $i = 86.73^\circ$, which is in
very good agreement with values cited by these authors.

The derived limb-darkening coefficients are in good agreement with the
ATLAS atmospheric models, combined with evolutionary stellar models,
over a stellar masses range within 1.05 to 1.25 $M_{\odot}$. There is
a minor difference between the stellar radii which best fit the
transit light curve ($1.20\pm0.02R_\odot$ for $M_* =1.1M_\odot$), and
the radii which produce the limb-darkenings best describing the ATLAS
models ($1.17\pm0.03 R_\odot$). As the differences are within the
error bars, we can not distinguish if they are caused by the errors
inherent in the observational data, or by the calibration of the ATLAS
models. More transit observations of \astrobj{HD~209458} with the same
instrument are planned for the next year. It will then be possible to
co-add light curves from individual transits, giving rise to an
improvement in the precision of the parameters derived in this paper,
especially with regard to the limb-darkening.

With the discovery of further transiting systems, and with the
availability of very high precision data from space missions, we
expect that transit observation will become an important tool in the
verification of model atmospheres, and will be able to constrain
stellar structure models over a wide range of stellar types.

\section{Acknowledgments}
We thank D. Charbonneau and T. Brown to forward us some of the
parameters they derived from observations with the Hubble Space
Telescope, and previous to publication. We thank the anonymous referee
for his comments, and both him and the Editor for their rapid
responses. The Observatorio de Sierra Nevada (OSN) is operated by the
Instituto de Astrof\'\i sica de Andaluc\'\i a.

\bibliography{HJDmain}

\begin{thebibliography}{13}
\expandafter\ifx\csname natexlab\endcsname\relax\def\natexlab#1{#1}\fi
\expandafter\ifx\csname url\endcsname\relax
  \def\url#1{{\tt #1}}\fi

\bibitem[{Brown} et~al.(2001){Brown}, {Charbonneau}, {Gilliland}, {Noyes}, and
  {Burrows}]{bcg+2001}
T.~{Brown}, D.~{Charbonneau}, R.~L. {Gilliland}, R.~W. {Noyes}, and
  A.~{Burrows}.
\newblock tbd.
\newblock {\em ApJL}, submitted, 2001.

\bibitem[{Castellano} et~al.(2000){Castellano}, {Jenkins}, {Trilling}, {Doyle},
  and {Koch}]{2000ApJ...532L..51C}
T.~{Castellano}, J.~{Jenkins}, D.~E. {Trilling}, L.~{Doyle}, and D.~{Koch}.
\newblock Detection of planetary transits of the star HD 209458 in the
  Hipparcos data set.
\newblock {\em ApJL}, 532:\penalty0 L51--L53, Mar. 2000.

\bibitem[{Charbonneau} et~al.(2000){Charbonneau}, {Brown}, {Latham}, and
  {Mayor}]{2000ApJ...529L..45C}
D.~{Charbonneau}, T.~M. {Brown}, D.~W. {Latham}, and M.~{Mayor}.
\newblock Detection of planetary transits across a sun-like star.
\newblock {\em ApJL}, 529:\penalty0 L45--L48, Jan. 2000.

\bibitem[{Claret}(1995)]{1995A&AS..109..441C}
A.~{Claret}.
\newblock Stellar models for a wide range of initial chemical compositions
  until helium burning. I. From x=0.60 to x=0.80 for z=0.02.
\newblock {\em A\&AS}, 109:\penalty0 441--446, Mar. 1995.

\bibitem[{Claret}(1998{\natexlab{a}})]{1998A&AS..131..395C}
A.~{Claret}.
\newblock Comprehensive tables for the interpretation and modeling of the light
  curves of eclipsing binaries.
\newblock {\em A\&AS}, 131:\penalty0 395--400, Sept. 1998{\natexlab{a}}.

\bibitem[{Claret}(1998{\natexlab{b}})]{1998A&A...335..647C}
A.~{Claret}.
\newblock Very low mass stars: non-linearity of the limb-darkening laws.
\newblock {\em A\&A}, 335:\penalty0 647--653, July 1998{\natexlab{b}}.

\bibitem[{Gronbech} et~al.(1976){Gronbech}, {Olsen}, and
  {Stromgren}]{1976A&AS...26..155G}
B.~{Gronbech}, E.~H. {Olsen}, and B.~{Stromgren}.
\newblock Standard stars uvby photoelectric photometry south of declination
  +10.
\newblock {\em A\&AS}, 26:\penalty0 155--176, Oct. 1976.

\bibitem[{Henry} et~al.(2000){Henry}, {Marcy}, {Butler}, and
  {Vogt}]{2000ApJ...529L..41H}
G.~W. {Henry}, G.~W. {Marcy}, R.~P. {Butler}, and S.~S. {Vogt}.
\newblock A transiting ``51 Peg-like'' planet.
\newblock {\em ApJL}, 529:\penalty0 L41--L44, Jan. 2000.

\bibitem[{Hestroffer} and {Magnan}(1998)]{1998A&A...333..338H}
D.~{Hestroffer} and C.~{Magnan}.
\newblock Wavelength dependency of the solar limb darkening.
\newblock {\em A\&A}, 333:\penalty0 338--342, May 1998.

\bibitem[{Jha} et~al.(2000){Jha}, {Charbonneau}, {Garnavich}, {Sullivan},
  {Sullivan}, {Brown}, and {Tonry}]{2000ApJ...540L..45J}
S.~{Jha}, D.~{Charbonneau}, P.~M. {Garnavich}, D.~J. {Sullivan}, T.~{Sullivan},
  T.~M. {Brown}, and J.~L. {Tonry}.
\newblock Multicolor observations of a planetary transit of HD 209458.
\newblock {\em ApJL}, 540:\penalty0 L45--L48, Sept. 2000.

\bibitem[{Mazeh} et~al.(2000){Mazeh}, {Naef}, {Torres}, {Latham}, {Mayor},
  {Beuzit}, {Brown}, {Buchhave}, {Burnet}, {Carney}, {Charbonneau}, {Drukier},
  {Laird}, {Pepe}, {Perrier}, {Queloz}, {Santos}, {Sivan}, {Udry}, and
  {Zucker}]{2000ApJ...532L..55M}
T.~{Mazeh}, D.~{Naef}, G.~{Torres}, D.~W. {Latham}, M.~{Mayor}, J.~{Beuzit},
  T.~M. {Brown}, L.~{Buchhave}, M.~{Burnet}, B.~W. {Carney}, D.~{Charbonneau},
  G.~A. {Drukier}, J.~B. {Laird}, F.~{Pepe}, C.~{Perrier}, D.~{Queloz}, N.~C.
  {Santos}, J.~{Sivan}, S.~. {Udry}, and S.~{Zucker}.
\newblock The spectroscopic orbit of the planetary companion transiting HD
  209458.
\newblock {\em ApJL}, 532:\penalty0 L55--L58, Mar. 2000.

\bibitem[{Robichon} and {Arenou}(2000)]{2000A&A...355..295R}
N.~{Robichon} and F.~{Arenou}.
\newblock HD 209458 planetary transits from Hipparcos photometry.
\newblock {\em A\&A}, 355:\penalty0 295--298, Mar. 2000.

\bibitem[{Soderhjelm}(1999)]{1999IBVS.4816....1S}
S.~{Soderhjelm}.
\newblock Possible detection of the planet transits of HD 209458 in the
  Hipparcos photometry.
\newblock {\em Informational Bulletin on Variable Stars}, 4816:\penalty0 1+,
  Dec. 1999.

\end{thebibliography}

\end{document}